\documentclass[aps,pra,groupedaddress,twocolumn]{revtex4-2}

\usepackage{amsmath}
\usepackage{amssymb,bm,graphicx}
\usepackage{subcaption}
\usepackage{multirow} 
\usepackage{color}

\begin{document}


\title{Robust Quantum Control for Bragg Pulse Design in Atom Interferometry}


\author{Luke S. Baker$^1$, Andre L. P. de Lima$^{1,2}$, Andrew Harter$^1$, Ceren Uzun$^1$, Liam P. Keeley$^3$, Jr-Shin Li$^2$, Anatoly Zlotnik$^1$, Michael J. Martin$^1$, Malcolm G. Boshier$^1$}
\affiliation{$^1$Los Alamos National Laboratory, $^2$Washington University in St. Louis, $^3$University of Colorado - Boulder}


\date{\today}

\begin{abstract}
We formulate a robust optimal control algorithm to synthesize minimum energy pulses that can transfer a cold atom system into various momentum states. The algorithm uses adaptive linearization of the evolution operator and sequential quadratic programming to iterate the control towards a minimum energy pulse that achieves optimal target state fidelity. Robustness to parameter variation is achieved using Legendre polynomial approximation over the domain of variation.  The method is applied to optimize the Bragg beamsplitting operation in ultra-cold atom interferometry. Even in the presence of 10-40\% variability in the initial momentum dispersion of the atomic cloud and the intensity of the optical pulse, the algorithm reliably converges to a control protocol that robustly achieves unprecedented momentum levels with high fidelity for a single-frequency multi-photon Bragg diffraction scheme (e.g. $|\pm 40\hbar k\rangle$).  We show the advantages of our method by comparison to stochastic optimization using sampled parameter values, provide detailed sensitivity analyses, and performance of the designed pulses is verified in laboratory experiments.
\end{abstract}


\maketitle

\section{Introduction}

Algorithms for optimal quantum control have been advanced and generalized during the past decades, and enable the current emergence of high-precision quantum circuits and sensing architectures \cite{mabuchi2005principles,dong2010quantum}.  
Established algorithms such as Gradient Ascent Pulse Engineering (GRAPE) \cite{khaneja2005optimal} iteratively adjust a control pulse towards one that maximizes the likelihood of realizing a target state with optimal fidelity.  Ensuring that the control is resilient to noise and platform disturbances within a design region is critical but challenging for practical quantum-enhanced sensing routines \cite{degen2017quantum, kosut2013robust, propson2022robust}.  While stochastic sampling has been shown to improve resilience \cite{wu2019learning}, a robust control approach that provides guarantees over a wide range of disturbances is highly desirable conceptually and computationally.  Here we address this long-standing algorithmic challenge and, crucially, demonstrate its feasibility in the laboratory.

Bose-Einstein condensates (BEC) have a rich history, and have been used in the metrology of fundamental constants with interferometers that function as highly sensitive probes of external interactions and inertial effects.  Given a flexible magnetic trap, an arbitrarily good vacuum, a true ground state, and nonpathological collisional properties, almost any magnetically trappable species can be evaporated to BEC \cite{cornell2002nobel}. Light-matter interactions can be used in atom interferometry, which can enable measurement that is theoretically much more precise than its optical counterpart.  For example, the fundamental limit on the signal-to-noise ratio of an atom Sagnac gyroscope is a factor of $10^{11}$ greater than an optical system with comparable enclosed areas and particle flux \cite{wang2005atom}. Light-atom interferometers use optical beam dividers based on the recoil effect experienced by an atom that absorbs a resonant photon \cite{baudon1999atomic}.  The atomic Sagnac effect was demonstrated using a beam of metastable Mg atoms by setting the apparatus on a rotatable platform \cite{riehle1991optical}, where instead of sending the atoms through successive laser beams, they are illuminated by successive coherent optical pulses.  Subsequent experiments have characterized the recoil frequency interferometrically, e.g. using a two optical standing wave pulses separated by a delay \cite{campbell2005photon}, and this has led to further exploration of how the momentum distribution of a BEC can be engineered by applying a sequence of standing-wave laser pulses \cite{edwards2010momentum}.  Quantum sensing has subsequently become a distinct and growing branch of research that aims to utilize the temporal and frequency interactions of spin qubits, trapped ions, and flux qubits with magnetic and electric fields, rotations, temperature, and pressure to exploit their strong sensitivity to external disturbances \cite{degen2017quantum}.

Concurrently, a rich variety of mathematical methods and algorithms have been developed for modeling and control of quantum dynamics \cite{dalessandro2021introduction}, which are used to address several broad classes of models and problems arising in quantum sensing and metrology \cite{mabuchi2005principles, altafini2012modeling}.  Canonical properties of controllability and observability of quantum mechanical systems were addressed decades ago \cite{huang1983controllability,belavkin1983theory}, and the theory of systems with continuous spectra was subsequently developed \cite{tarn2000controllability}.  Later studies examined existence of optimal controls for fixed-endpoint problems for the Schroedinger equation and developed numerical methods for directly controlling the wave packet \cite{peirce1988optimal}, as well as quadratic optimal control of two-level systems in Hilbert space \cite{dalessandro2001optimal}.  Other studies applied Lyapunov-based methods and continuous state estimation and filtering schemes \cite{hou2012optimal,doherty1999feedback}, as well as sliding-mode control \cite{dong2009sliding}.  We refer the reader to a concise review that introduces quantum states, measurements, control models, controllability, Lyapunov-based design, and coherent feedback control \cite{dong2010quantum}, as well as a subsequent update with more recent literature \cite{dong2022quantum}. 

The importance of control concepts for practical shaping of pulse sequences to achieve selective transformations in the degrees of freedom of atoms, molecules, or materials has long been recognized \cite{walmsley2003quantum}.  The ability to achieve robustness in nuclear magnetic resonance measurements is significantly enabled by advanced signal processing and control-theoretic techniques \cite{levitt1986composite, pauly1991parameter,khaneja2005optimal,li2011optimal}.  Among the many theoretically-inspired approaches to manipulate quantum systems, such as pulse-timing control, coherent control, stimulated-Raman-adiapatic-passage (STIRAP), and others, a consensus has emerged that optimal control methods developed in engineering settings are the most powerful \cite{goswami2003optical,werschnik2007quantum}.  The advantages in precision and state fidelity come at the cost of increased complexity in control pulses and complication of associated synthesis methods \cite{mizrahi2014quantum,peterson2020enhancing,wu2019learning,ge2021risk,petersson2025time,oneil2024robustness}.  Significant attention has therefore been applied to characterize and improve the performance of GRAPE \cite{khaneja2005optimal,lu2024optimal,bhole2018practical,defouquieres2011second}. Precise and reliable fidelity is particularly important for realizing ultra-sensitive coherent control processes such as optically-detected magnetic resonance (ODMR) or spectroscopy of single molecules \cite{barry2020sensitivity,hirose2016coherent}.  Such methods require coordinated application of multi-scale electromagnetic and optical fields to realize complex quantum logic for manipulation and measurement \cite{lovchinsky2016nuclear}.

The design of quantum control protocols that are insensitive to uncertainty and/or variability in  intrinsic system parameters and applied fields is a broad research area called robust quantum control \cite{weidner2025robust,dong2023learning,koswara2021robust}.  Early methods primarily addressed open-loop pulse designs with performance that is insensitive to a range of values for a static parameter \cite{daems2013robust}, by extending GRAPE using sampling or more sophisticated spectral methods \cite{ruths2011multidimensional,li2011optimal}.  A comprehensive exposition of iterative methods for robust control using sequential convex programming (SCP) provides a literature review, problem formulations, nomenclature and parametrizations, and includes computational examples of pulse synthesis for a one-qubit gate \cite{kosut2013robust}.  Contemporary techniques can provide robustness guarantees with respect to both static and time-dependent parameter uncertainty \cite{propson2022robust}. 
Techniques that result in smooth pulses are often sought to avoid degradation of fidelity in practice due to low-pass filtering effects \cite{barnes2015robust,li2011optimal}, although smoothness comes at the cost of design complexity \cite{koswara2021quantum,gungordu2019analytically}.  Various learning, data-driven, and probabilistic approaches have been proposed \cite{wu2019learning,ge2021risk,oneil2024robustness}, and practical techniques in magnetometry \cite{clevenson2018robust, oshnik2022robust}, dynamical decoupling \cite{shim2012robust}, spin polarization \cite{schwartz2018robust}, quantum logic gates \cite{george2025minimal,zhang2023robust}, and Bragg atom interferometry \cite{louie2023robust} have been augmented with robustness properties.  A compelling emerging application considers the suppression of crosstalk in bosonic quantum logic gates, which refers to interference due to multiple modes coupled to the same ancilla \cite{you2024crosstalk}.  Recent methods also consider robust analogues of Floquet and Hamiltonian engineering \cite{zhou2023robust,boyers2019floquet}.  

The theory of matter-wave interferometry is over a century old \cite{bouyer2014centenary} and the various regimes of light diffraction by periodic phase gratings are well-characterized \cite{moharam1978criterion}.  The basic proof-of-principle demonstration of using light scattered by trapped atoms for measurement was demonstrated decades ago \cite{giltner1995atom}.  Research in quantum metrology has experienced significant subsequent growth, with experiments on the use of trapped atoms interrogated by optical fields as interferometers to precisely measure rotations, gravitational fields, and multi-axis inertial changes \cite{gustavson1997precision,peters2001high, dickerson2013multiaxis}. Particular attention has been applied to designs that are compact and robust to disturbances and noise \cite{menoret2018gravity,fang2012advances}.  The various matter-wave diffraction regimes and phase responses to optical interrogation have been explored \cite{muller2008atom,decamps2018phase}, and numerous hardware designs, control systems, and measurement techniques have been developed \cite{sakaguchi2016matter,hinton2017portable,hinton2017portable, lee2022compact,patil2020system,krzyzanowska2023matter}.  The sensitivity of metrology methods that use matter-wave splitting fundamentally depends on the path length of two optical beams before they are coupled out of an interferometer, and maximizing this path length for a given system depends on the performance of control techniques used to excite a BEC to a high momentum level with high fidelity \cite{wu2005splitting,wu2004high,jager2014optimal}.

Though the first atom interferometry demonstration experiments were performed decades ago, matter-wave sensors are still in process of moving from the laboratory into the field \cite{bongs2019taking,narducci2022advances}.  Many studies in just the past several years have sought to augment the latest atomic microtrap technology with advanced control synthesis methods \cite{templier2022tracking,saywell2023enhancing,wang2024robust}.  Recent studies have also sought pulse sequences that are robust to parameter variation or noise, and popular approaches use data-driven methods or machine learning \cite{louie2023robust,ledesma2024demonstration,colussi2024machine}.

In this study, we introduce a powerful optimal pulse synthesis algorithm with improved performance and resilience for high momentum beam-splitting in a matter-wave interferometer.  Our robust optimal quantum control method uses sequential quadratic programming (SQP), similar to an earlier SCP approach \cite{kosut2013robust}, that incorporates a spectral approximation method over the domain of parameter variation \cite{li2022moment} to provide robustness guarantees.  The methodology is developed for a two-dimensional Sagnac atom interferometer in which Bose-condensed $^{87}$Rb atoms propagate within a waveguide formed by a collimated laser beam, in a matter-wave analog of the fiber-optic gyroscope \cite{krzyzanowska2023matter}.  The algorithm is based on our preliminary study on general closed quantum systems \cite{baker2024convergence}, which was previously demonstrated using numerical simulations involving the Raman-Nath equations \cite{delima2023optimal}.  The key advances that we describe in this paper are (1) unprecedented momentum levels, e.g., $|\pm 40\hbar k\rangle$, compared to the state of the art (i.e., $|\pm 8\hbar k\rangle$) in a single-frequency multi-photon Bragg diffraction scheme \cite{cassidy2021improved}; (2) insensitivity to variation in photon recoil frequency and initial momentum distribution; and (3) experimental validation to prove that complicated pulse sequences can be successfully implemented to achieve high-fidelity momentum state preparation.  

The rest of the manuscript is organized as follows.  Section \ref{sec:evolution} derives a Legendre polynomial series representation of a Hamiltonian with dependence on continuous parameters that is used to represent the evolution of a quantum ensemble system subject to external control fields.  Section \ref{sec:control} formulates two optimization problems in function space that are iteratively solved in two stages to synthesize a fixed endpoint minimum energy optimal control subject to restrictions on amplitude or other signal properties.  The first stage ensures high fidelity in the prepared state, and the second stage minimizes the total energy in the synthesized pulse.  In Section \ref{sec:bragg_analysis} we describe the application of our method for robust Bragg pulse design in atom interferometry, including detailed analyses of algorithm performance in simulation.  Section \ref{sec:results} shows the results of simulations to validate robust Bragg beamsplitting pulse synthesis, as well as the outcomes of several laboratory experiments that verify high momentum state beamsplitting in practice.  Section \ref{sec:conc} concludes with discussions and ideas for future work.

\section{Evolution of a Quantum Ensemble} \label{sec:evolution}

Consider a closed quantum system over a finite-dimensional Hilbert space in which the wave vector evolves according to the Schr{\"o}dinger equation
\begin{equation} \label{eq:schrodinger}
    i\hbar \frac{d |\Psi(t)\rangle}{dt}=\left( \sum_{m=1}^M  \gamma_m u_m(t) \hat H_m \right)|\Psi(t)\rangle,
\end{equation}
where $\hat H_m$ represents an individual matrix component of the Hamiltonian. The evolution of the state is influenced by control variables $u_m(t)$ for $m=1,\ldots,M$ that represent the amplitudes of optical or electromagnetic fields applied to the system.  Static components of the Hamiltonian are described by defining the corresponding control as unity for all time.  Each parameter $\gamma_m$ may assume any value in a specified closed interval denoted by $[\gamma_m^{\min},\gamma_m^{\max}]$.  Writing the vector of parameters as $\bm \gamma = (\gamma_1, \dots, \gamma_M)$, we denote the solution to equation \eqref{eq:schrodinger} for a given value of $\bm \gamma$ by $|\Psi(t,\bm \gamma) \rangle$.  The parameter vector $\bm \gamma$ represents the effects of miscalibration and platform noise that may appear as parameter variability within the components of the Hamiltonian or the amplitudes of the applied fields.  For ease of exposition, we follow previous approaches and assume that the Hamiltonian depends linearly on parameters $\gamma_m$ \cite{khaneja2005optimal}.  The above formulation warrants the design of control pulses that are robust or insensitive to coherent errors and can be used to model a broad range of desirable applications in atom interferometry \cite{wang2005atom, saywell2023enhancing, louie2023robust, colussi2024machine}, quantum computing \cite{divincenzo1995quantum}, and nuclear magnetic resonance spectroscopy \cite{morris1992diffusion}. 

\vspace{-2ex}
\subsection{Wavefunction Expansion on Parameter Domain} \label{subsec:expansion}

We suppose that a parameterized Hamiltonian governs the evolution of a finite-dimensional wave vector dependent on a continuum of parameter values.  A common approach for reduction to a finite-dimensional representation is based on direct \cite{khaneja2005optimal} or stochastic \cite{wu2019learning, ge2021risk} sampling over design intervals for each varying parameter.  We use an alternative approach similar to the moment quantization method in which dependence of a finite state vector on a parameter is represented by a product of Legendre polynomials \cite{li2022moment}. Because of exponential decay properties of Legendre expansions \cite{wang2012convergence}, the method of moments and similar spectral approximations more compactly and accurately represent parametrized dynamic constraints in optimization problems.  This enables faster convergence of optimization algorithms than representations based on sampling, and this advantage will be demonstrated in the examples below.

Introducing multi-index notation, we denote an ordered array of $M$ integers by $\bm n=(n_1,\dots,n_M)$ corresponding to the vector of parameters $\bm \gamma = (\gamma_1, \dots, \gamma_M)$ that take values on the $M$-dimensional domain   
$\mathcal D = [\gamma_1^{\min},\gamma_1^{\max}]\times[\gamma_2^{\min},\gamma_2^{\max}]\times\cdots\times[\gamma_M^{\min},\gamma_M^{\max}]$.  We consider a Legendre expansion of degree $N_m$ for each parameter $\gamma_m$ varying on $[\gamma_m^{\min},\gamma_m^{\max}]$, so that the basis elements in the scalar product space of Legendre polynomials over the domain $\mathcal D$ are $\ell_{\bm n}(\bm \gamma)=\ell_{n_1}(\gamma_{1})\cdots \ell_{n_M}(\gamma_{M})$ for $\bm n \in \mathcal N$. Here, $\ell_n(\gamma_m)$ represents the normalized Legendre polynomial of degree $n$ with the domain shifted to the interval $[\gamma_m^{\min},\gamma_m^{\max}]$. The multi-index $\bm n$ thus takes values in the set $\mathcal N=[N_1]\times[N_2]\times\cdots\times[N_M]$, where we denote $[n]=\{0,1,\ldots,n\}$ as the set of the first $n+1$ non-negative integers.  Let us denote an enumeration of the multi-index values by $\sigma:|\mathcal N|\to\mathcal N$.

We suppose that the parametrized wavefunction solution $|\Psi(t,\bm \gamma) \rangle$ lies on a separable Hilbert space $\mathcal L$ on $\mathcal D$.  By truncating the Legendre expansion to polynomials of sufficiently high degrees, we obtain an approximate representation of the parametrized wavefunction $|\Psi(t,\bm \gamma) \rangle$, using basis elements indexed by $\bm n\in \mathcal N$, of the form
\begin{equation} \label{eq:Legendre_expansion}
     |\Psi_{\mathcal N}(t,\bm \gamma) \rangle=\sum_{\bm n\in\mathcal N}| \psi_{\bm n}(t)\rangle \ell_{\bm n}(\bm \gamma).
\end{equation}
In the above equation, the time-dependent \emph{wavefunction moments} are defined by the formula
\begin{equation} \label{eq:Legendre_coefficients}
    |\psi_{\bm n}(t)\rangle=\int_{\mathcal D} |\Psi(t,\bm \gamma)\rangle \ell_{\bm n}(\bm \gamma)d\bm \gamma.
\end{equation}
Geometrically, the wavefunction moment $|\psi_{\bm n}(t)\rangle$ represents the projection of the parametrized wavefunction $|\Psi(t,\bm \gamma)\rangle$ onto the subspace of $\mathcal L$ spanned by the Legendre basis element $\ell_{\bm n}(\bm \gamma)$.  Whereas the moment quantization approach for semiclassical models \cite{li2022moment} results in a Legendre basis expansion for which the moments are vector-valued, our approach of applying moment quantization to the Schr\"odinger equation \eqref{eq:schrodinger} directly results in moments $|\psi_{\bm n}(t)\rangle$ that are wavefunctions.  In the following, we may drop the dependence on time and write $|\psi_{\bm n}\rangle$ for ease of exposition.
\vspace{2ex}

\vspace{-2ex}
\subsection{Wavefunction Moment Dynamics} \label{subsec:legendre}

The evolution equations for the wavefunction moments $|\psi_{\bm n}\rangle$ for $\bm n\in\mathcal N$ are obtained by differentiating Eq. \eqref{eq:Legendre_coefficients} with respect to time, interchanging the order of differentiation and integration, and substituting the expression of the derivative using Eq. \eqref{eq:schrodinger}.  By the orthogonality and recurrence relations of Legendre polynomials, the evolution of $|\psi_{\bm n}\rangle$ for a given $\bm n\in\mathcal N$ is governed by \cite{baker2024convergence, baker2025robust}
\begin{align} 
    i\hbar \frac{d|\psi_{\bm n}\rangle}{dt} &= \sum_{m=1}^M c_{n_m-1}\underline{\gamma}_m |\psi_{\bm n_m^-}\rangle \qquad \nonumber \\
     & \qquad + \sum_{m=1}^M c_{n_m}\overline{\gamma}_m |\psi_{\bm n}\rangle \qquad \nonumber \\
    & \qquad + \sum_{m=1}^M c_{n_m+1}\underline{\gamma}_m |\psi_{\bm n_m^+}\rangle, \label{eq:robust_legendre}
\end{align}
where $ \bm n_m^{\pm} = (n_1,\dots, n_m\pm 1, \dots, n_M)$ denotes increment or decrement of the $m$-th entry of multi-index $\bm n$, and
\begin{equation} 
 c_{n} = \frac{n+1}{\sqrt{(2n+3)(2n+1)}}. \label{eq:legendre_recurrence}
\end{equation}
The parameters $\overline \gamma_m=(\gamma_m^{\max}+ \gamma_m^{\min})/2$ and $\underline \gamma_m=(\gamma_m^{\max}- \gamma_m^{\min})/2$ represent the midpoint and maximum absolute variation of the parameter $\gamma_m$ over its associated design interval.
Equation \eqref{eq:robust_legendre} indicates that the dynamics of Legendre basis coefficients are coupled to those of nearest degree polynomials in Legendre space.  We denote the tensor product of the wavefunction moments by
\begin{equation} 
|\psi(t)\rangle = \otimes_{\bm n \in \mathcal N} |\psi_{\bm n}(t)\rangle, \label{eq:wavefunction_embedding}
\end{equation}
where the tensor product is applied using the multi-index enumeration $\sigma$, i.e., $|\psi\rangle=|\psi_{\sigma(1)}\rangle \otimes |\psi_{\sigma(2)}\rangle \otimes \cdots \otimes |\psi_{\sigma(|\mathcal N|)}\rangle$.  Using this notation, the dynamics of all the wavefunction moments may be written in the matrix form
\begin{equation} \label{eq:robust_schrodinger}
    i\hbar \frac{d|\psi(t)\rangle}{dt}=\left(\sum_{m=1}^M u_m(t)H_{m} \right)|\psi(t)\rangle,
\end{equation}
where the spectral expansion of $\hat H_m$ over parameter space is defined by the chain of tensor products
\begin{equation}  \label{eq:robust_schrodinger_hamiltonian}
    H_m= \otimes_{j=0}^{m-1}I_{N_j+1} \otimes \Gamma_{m} \otimes_{j=m+1}^{M}I_{N_j+1} \otimes \hat H_m.
\end{equation}
The summation over nearest neighbors is described by the tridiagonal matrices
\begin{equation} \label{eq:robust_schrodinger_coeff}
    \Gamma_{m}=\begin{pmatrix}
        \overline \gamma_m & c_1\underline \gamma_m &  &  & &    \\
        c_1\underline \gamma_m & \overline \gamma_m & c_2\underline \gamma_m &  & &  \\
         & c_2\underline \gamma_m & \overline \gamma_m &  & &   \\
         & & & \ddots & &  \\
         & & & & \overline \gamma_m &  c_{N_m}\underline \gamma_m \\
         & & & & c_{N_m}\underline \gamma_m &  \overline \gamma_m
         \end{pmatrix}.
\end{equation}
If the expansion is restricted to zero degree polynomials by choosing $N_m=0$ for all $m$, then the spectral approximation in Eq. \eqref{eq:robust_schrodinger} reduces to a single evolution equation equivalent to sampling Eq. \eqref{eq:schrodinger} at the average value $\gamma_m=\overline \gamma_m$ for all $m$. Moreover, by directly sampling Eq. \eqref{eq:schrodinger} at $N_m+1$ chosen values over each interval $[\gamma_m^{\min},\gamma_m^{\max}]$ and arranging the sampled systems with the above ordering, we obtain Eq. \eqref{eq:robust_schrodinger} but with $\Gamma_m$ replaced by a diagonal matrix of the chosen samples.  We conclude that Eq. \eqref{eq:robust_schrodinger} may be used to model parameter variation by expansion over Legendre polynomials or direct sampling by modifying $\Gamma_m$, and therefore offers a general formulation for comparison of these two approaches.
Our subsequent analysis uses the spectral approximation in equations \eqref{eq:robust_schrodinger}-\eqref{eq:robust_schrodinger_coeff}.
\vspace{2ex}

\section{Quantum Optimal Control} \label{sec:control}

We formulate an optimal control algorithm to actuate a transfer of the dynamics \eqref{eq:robust_schrodinger}-\eqref{eq:robust_schrodinger_coeff} from a prepared initial state $|\psi_0\rangle$ to a specified target state $|\psi_T\rangle$ over a given time interval $[0,T]$. 
We assume for simplicity that the chosen time of application is discretized into $K$ equal steps of duration $\Delta t=T/K$. For constant control inputs during each step $[t_k,t_{k+1})$, the wavefunction in Eq. \eqref{eq:robust_schrodinger} evolves according to $|\psi_{k+1}\rangle=U_k|\psi_k\rangle$, where the unitary operators are defined by
\begin{equation} \label{eq:unitary_matrix}
    U_k=\text{exp} \left( -\frac{i \Delta t}{\hbar} \sum_{m=1}^M u_m(t_k)H_m\right).
\end{equation}
The subscript $k$ indicates the evaluation at time $t_k$ of the quantity to which it associates. 

Mathematical programming methods consider operations using finite vector representations, and here we suppose that the wavefunctions $|\psi_k\rangle$ can be adequately approximated using a finite basis $\{|e_j\rangle\}_{j=1}^N$ for all time points $k=1,\ldots,K$ according to
\begin{equation} \label{eq:basis_approximation}
    |\psi_k\rangle=\sum_{j=1}^Na_{jk}|e_j\rangle,
\end{equation}
so that the time-indexed vector of coefficients $\vec\psi_k = (a_{1k},\ldots,a_{Nk})'$ is used to represent $|\psi_k\rangle$ in the optimization algorithm.  In this section, we describe an optimization algorithm that is agnostic to the basis representation, assuming that an appropriate separable Hilbert space $\mathcal H$ for the wavefunction $|\psi_k\rangle$ exists.
We suppose that $\vec\psi_k$ is a basis coefficient representation of $|\psi_k\rangle$.  Successively applying single time-step transition operators corresponding to the selected controls results in a transfer to the terminal state
\begin{equation}  \label{eq:unitary_evolution}
    |\psi_K\rangle=U_{K-1}U_{K-2}\cdots U_1U_0 |\psi_0\rangle.
\end{equation}
The nonlinear evolution equation \eqref{eq:unitary_evolution} is adaptively linearized and used as a constraint in iterative quadratic programming in our control design algorithm.  The method has two stages, the first to minimize terminal state error and the second to minimize control energy while keeping the terminal state fixed.

\vspace{-2ex}
\subsection{Control Design for Transfer to Target State} \label{subsec:target}

For each $k$, define $\mathfrak u_k=(u_0(t_k), u_1(t_k),\dots,u_M(t_k))'$ and use the time sampled vector to construct the control vector $u=(\mathfrak u_0', \mathfrak u_1',\dots, \mathfrak u_{K-1}')'$.  Likewise, define the matrices 
\begin{equation}
    J_k=\left( \frac{\partial \vec\psi_K}{\partial u_0(t_k)},\frac{\partial \vec\psi_K}{\partial u_1(t_k)},\dots, \frac{\partial \vec\psi_K}{\partial u_M(t_k)} \right),
\end{equation}
and construct the Jacobian $J=(J_0,J_1,\dots,J_{K-1})$. Given a control $ u$ and the corresponding terminal state $|\psi_K\rangle$ computed using Eq. \eqref{eq:unitary_evolution}, the control algorithm seeks to update the control $u \rightarrow  u+\delta u$ in such a way that the update $\vec\psi_K\rightarrow \vec\psi_K+\delta \vec\psi_K$ of the vector of basis coefficients for the terminal state improves the optimization objective.  Although other objectives may be formulated, we consider minimization of the terminal error $\|\vec\psi_K-\vec\psi_T\|^2=(\vec\psi_K-\vec\psi_T)'(\vec\psi_K-\vec\psi_T)\approx 1 - |\langle \psi_K | \psi_T \rangle|^2$, which has advantages for constrained optimization with vector space methods \cite{luenberger1997optimization}.  To first order, the updates are related by the linear equation
\begin{eqnarray} \label{eq:linear_transition}
    \delta \vec\psi_K=J\delta  u.
\end{eqnarray}
Because this iterative updating is standard for gradient-based methods, the simulation in Eq. \eqref{eq:unitary_evolution} and the Jacobian $J$ may be computed by existing methods \cite{georgescu2014quantum, de2011second}.

As an alternative to gradient updates, $\delta u$ is defined to be the solution of the quadratic program given by
\begin{equation} \label{eq:QP_error}
\begin{array}{ll}
\text{minimize:}   & \|J\delta u + \vec\psi_K-\vec\psi_T\|^2+\lambda \|\delta u \|^2 \\
\text{subject to:}   & \text{dynamic constraints \eqref{eq:unitary_matrix}-\eqref{eq:unitary_evolution}}, \\
&\text{control restrictions}.
 \end{array}
\end{equation}
The vector $\delta u$ is the variation that improves the terminal fidelity for transfer to the target state achieved by the control $u$ at the current iterate.
An advantage of this formulation is that it directly enables the user to define control restrictions in the form of inequality constraints. By the term `control restrictions', we refer to specified lower and upper bounds on any finite number of linear combinations of the components of the control vector. These restrictions are considered to be a practical feature of the proposed algorithm because they can ensure an engineered outcome in which the intensity, spectrum, and potentially total energy of the prepared control may be directly restricted to comply with equipment operating limits. A positive regulation parameter $\lambda$ is introduced into the objective to restrict the size of the update between iterations.  Choosing sufficiently large $\lambda$ enables regulation of  the local dynamics and hence the accuracy of linearization to arbitrary precision.  The algorithm is initiated by choosing a random control $u$ that satisfies the control restrictions and a large enough value for $\lambda$ that is afterward decreased as a function of iteration count if the difference between two successive objective values falls below a given threshold. The algorithm terminates whenever the error reaches a given tolerance.

\begin{figure*}
    \centering
        \includegraphics[width=\linewidth]{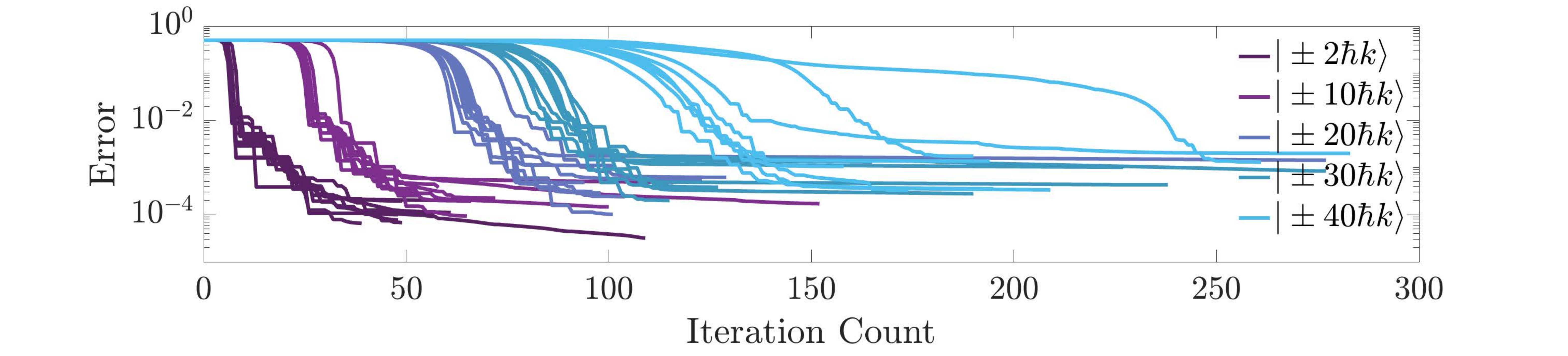} 
                \vspace{-2ex}
    \caption{Convergence of the iterative algorithm for selected target momenta $|\pm 2n_0 \hbar k\rangle$.}
    \label{fig:error}
\end{figure*}

\vspace{-2ex}
\subsection{Control Attenuation to Minimal Energy} \label{subsec:energy}

A control of minimum energy is usually synthesized by introducing penalty terms to the objective function in Eq. \eqref{eq:QP_error}.  The energy refers to the $L_2$-norm of the control magnitude, and is sometimes referred to as the fluence in the quantum control literature. the sometimes referred to as  Because this approach could lead to competition between control energy and terminal state fidelity, we resolve this trade-off by formulating a minimum energy control algorithm that is applied in stages to first maximize terminal state fidelity by way of problem \eqref{eq:QP_error} and then minimize control energy while achieving the same fidelity.  For the latter, the control pulse is gradually adjusted by solving for the variation $\delta u$ by subsequent iterations of the quadratic program \cite{vu2023iterative}
\begin{equation} \label{eq:QP_energy}
\begin{array}{ll}
\text{minimize:}    & \|u+\delta u \|^2 +\mu \|\delta u\|^2, \\
\text{subject to:}    & PJ\delta u = 0, \\
                & \text{dynamic constraints \eqref{eq:unitary_matrix}-\eqref{eq:unitary_evolution}}, \\
              &\text{control restrictions},
 \end{array}
\end{equation}
where $\mu$ serves the same purpose as $\lambda$ does in the formulation in Eq. \eqref{eq:QP_error}. The iteration terminates when $\delta u \equiv 0$ minimizes problem \eqref{eq:QP_energy}, and the control $u$ of minimum energy $\|u\|^2$ is reached.  Using Eq. \eqref{eq:linear_transition}, the equality constraint $PJ\delta u=0$ ensures that the terminal state fidelity achieved by solving the robust optimal state transfer formulation in Eq. \eqref{eq:QP_error} is preserved as the minimum energy algorithm to solve the formulation in Eq. \eqref{eq:QP_energy} progresses through its iterations.  In this way, fidelity is not affected by minimizing control energy. The matrix $P$ represents a projection onto the space orthogonal to the target state to provide the minimum energy algorithm with flexibility to alter the phase of the target state for fixed magnitude in probability. Although the robust state transfer and minimum energy algorithms are presented individually, we note that they may be integrated into one algorithm that solves both quadratic programs \eqref{eq:QP_error} and \eqref{eq:QP_energy} in succession with each iteration.  Another approach is to alternate between formulations \eqref{eq:QP_error} and \eqref{eq:QP_energy} and apply an adaptive number of iterations before switching.  Such algorithmic extensions are beyond the scope of this study.  For the interested reader, the code used to implement our algorithm for this study is available in open source \cite{baker2025beast}.

\vspace{3ex}
\section{Atom Interferometry} \label{sec:atom_interferometry}
\vspace{-1ex}

The robust quantum control technique described in Section \ref{sec:control} is applied to design Bragg pulses for symmetric beamsplitters in ultra-cold atom interferometry.  Beamsplitters often rely on multi-photon Bragg scattering, where counter-propagating laser beams form a standing-wave that coherently transfers momentum to atoms and splits their wave packets into distinct momentum states \cite{wu2004high}. This process forms the foundation of atom interferometry by creating the necessary path separation for interference measurements. Recent efforts utilize rectangular, triangular, and Gaussian pulse sequences \cite{wang2005atom, wu2005splitting,edwards2010momentum, muller2008atom, cassidy2021improved}, in which the magnitude and duration of pulses are typically treated as optimization variables.  While the small number of parameters in these methods make them appealing, predefined signal shapes can severely limit the space over which the optimizer seeks a solution and result in sub-optimality and low fidelity.  To improve fidelity, robust optimal control has recently been applied to design beamsplitting pulses \cite{saywell2023enhancing, delima2023optimal, baker2024convergence}, with which experimental advantage over some contemporary methods has been demonstrated \cite{saywell2023enhancing}. Design of robust pulses capable of realizing high momentum beamsplitters with high fidelity has remained an open challenge until now.  

\vspace{-2ex}
\subsection{Cold Atom in a Standing-wave Potential}  \label{subsec:atomtrap}

The evolution of a sample of ultra-cold atoms in a one-dimensional standing-wave potential is assumed to be governed by the Schr{\"o}dinger equation \cite{wu2005splitting, cassidy2021improved}
\begin{equation}\label{eq:bec}
    i \frac{\partial \Psi}{\partial t} = -\frac{\hbar}{2 m}\frac{\partial^2\Psi}{\partial x^2} +u(t)\cos(2k_0x)\Psi,
\end{equation}
where $u(t)$ is the amplitude of the light shift potential and $k_0$ is the vacuum wave number of the photons. Although additional control variables may potentially enable even higher fidelity, we assume control of only optical intensity and use this stage to highlight the capabilities of the algorithm even under such limited control. Eliminating the need for additional control variables reduces experimental complexity, which distinguishes our approach from that described in Refs. \cite{saywell2023enhancing, louie2023robust}, where parameters such as the relative detuning and phase of the counter-propagating beams are controlled. 
By expanding the wavefunction in the momentum basis as
\begin{equation} \label{eq:wavefunction}
\Psi(t,x)=\sum_{n}\int d\bm k C_{2n}(t,\bm k)e^{i(2nk_0+k)x}
\end{equation}
and substituting the superposition into Eq. \eqref{eq:bec}, the evolution of the coefficients may be expressed as \cite{wu2005splitting}
\begin{equation}\label{eq:Raman_Nath}
    i \dot C_{2n} \!=\! \omega_r\left( 2n + \frac{k}{k_0}\right)^{2}\!\!\!{C}_{2n} +\gamma\frac{u(t)}{2}\!\left({C}_{2n-2} \!+\! {C}_{2n+2}\right),
\end{equation}
where $\omega_r=\hbar k_{0}^2/2m$ is the photon recoil frequency.  The parameters $k/k_0$ and $\gamma$ are considered variable over respective ranges to compensate for variation of momentum across the atomic population and optical intensity, respectively. 

The set of scalar equations indexed by the momentum level $n$ is truncated with integral steps from $n=-N$ to $n=N$.  It will be clear from context whether $k$ and $n$ denote conventional quantities in atom interferometry or the above time step and Legendre polynomial indices. 
By adopting common physical assumptions \cite{delima2023optimal,cassidy2021improved}, the diffraction dynamics in equation \eqref{eq:Raman_Nath} may be approximated by the dynamics of the vector $\vec \psi=[C_0,C_2,\dots,C_{2N}]'$ of nonnegative momentum basis coefficients, which are
\begin{equation} \label{eq:bec_rn}
   i \frac{d}{dt}\vec \psi =\left(\gamma_1 u_1(t) \hat H_1+ \gamma_2 u_2(t) \hat H_2\right) \vec \psi,
\end{equation}
where the $(N+1)\times (N+1)$ matrices $\hat H_1$ and $\hat H_2$ are defined by

\begin{eqnarray}
   \!\!\!\!\!\!\!\!\!  \hat H_1= \begin{bmatrix}
        0 & 0 &  &     \\
       0  & 4 & 0 &   \\
         & 0 & \ddots & 0   \\
         & & 0 &  (2N)^2
    \end{bmatrix}\!\!, \,\,   \hat H_2= \frac{1}{2}\begin{bmatrix}
        0 & \sqrt{2} &  &     \\
       \sqrt{2}  & 0 & 1 &   \\
         & 1 & \ddots & 1   \\
         & & 1&  0 
    \end{bmatrix}\!\!. \,\, \label{eq:B0_matrix} 
\end{eqnarray}

\noindent The above dynamics are a finite-dimensional approximation of beamsplitter dynamics in the form of equation \eqref{eq:schrodinger}, with $u_1(t)\equiv 1$.
By specifying a target momentum level $n_0\ll N$, the controller seeks a robust pulse that probes the atomic cloud from the initially prepared zero momentum state $|0\hbar k\rangle$ to the excited state $|\pm 2n_0 \hbar k\rangle$, where $|2n\hbar k\rangle=C_{2n}$.  Following preliminary simulations, we find that representations that truncate the series at order $N=18+2n_0$ are sufficient to accurately model dynamics in the populated momentum states. This threshold is used for the simulations below.

\begin{figure*}
\vspace{1ex}
    \centering
        \includegraphics[width=.85\linewidth]{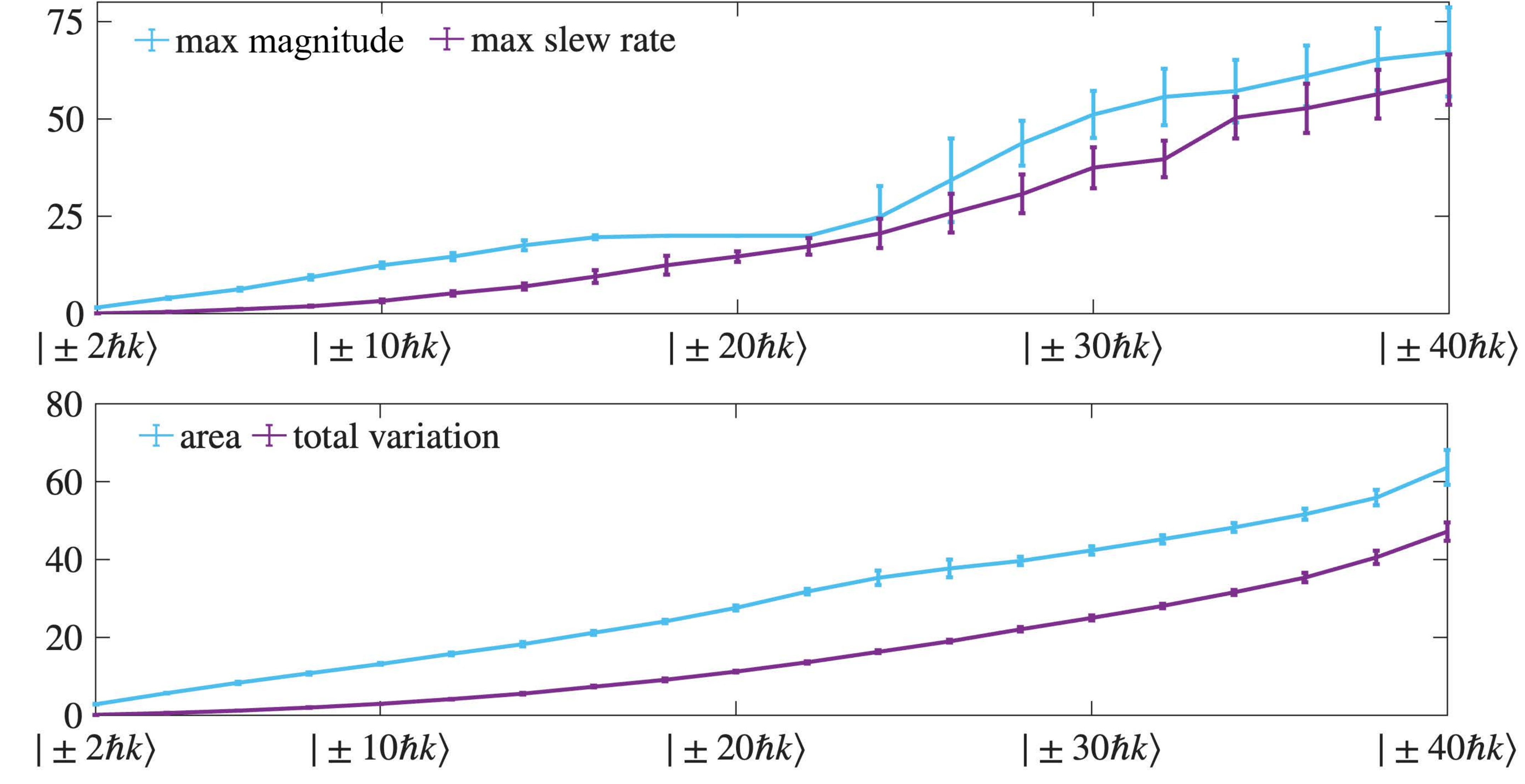}
    \caption{Metrics of maximum magnitude $MM[u]$, maximum slew rate $MS[u]$, area $A[u]$, and total variation $TV[u]$ of pulses $u$ for momentum states from $|\pm 2 \hbar k\rangle$ to $|\pm 40 \hbar k\rangle$.  These metrics are defined in Section \ref{sec:intensity}.}
    \vspace{-2ex}
    \label{fig:energy}
\end{figure*}

\begin{figure*}
    \centering
        \includegraphics[width=.95\linewidth]{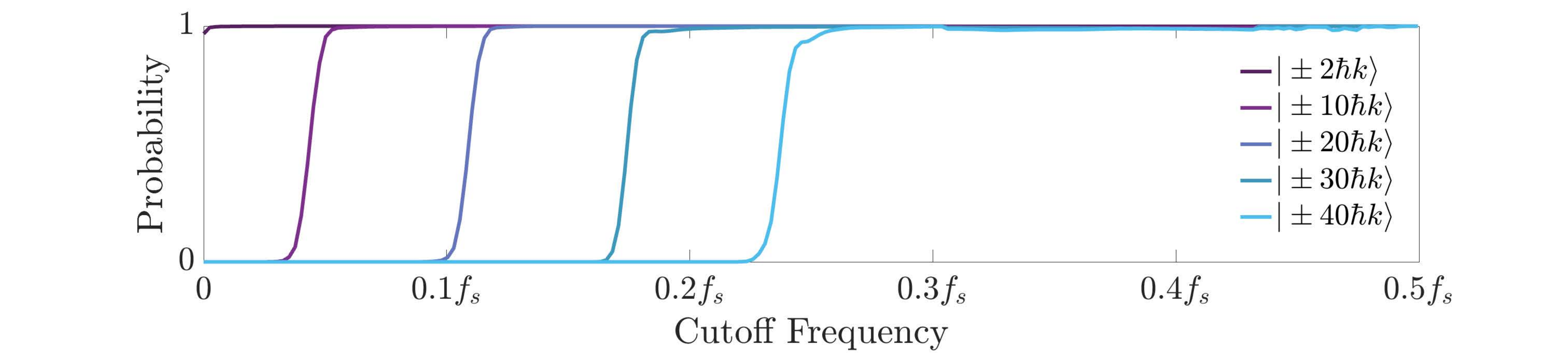}
    \caption{Probability of selected target momenta $|\pm 2n_0 \hbar k\rangle$ corresponding to IIR filtered control pulses defined by the cutoff frequency in terms of the sampling frequency $f_s$.}
    \vspace{-4ex}
    \label{fig:frequency}
    \vspace{2ex}
\end{figure*}

\section{Bragg Beamsplitter Analysis} \label{sec:bragg_analysis}

Several studies are presented to showcase various properties of the control algorithm applied to the design of Bragg beamsplitters. Control pulses are designed to populate symmetric momentum states ranging from $|\pm 2\hbar k\rangle$ to $|\pm 40\hbar k\rangle$.   Because high momentum beamsplitting challenges algorithmic convergence of both optimal control and conventional methods, we devise a novel application of the proposed algorithm and use it to demonstrate that realization of high momentum beamsplitters can be achieved with high fidelity.  Moreover, to demonstrate repeatability of the algorithm, ten realizations are computed for ten corresponding random initializations of the control pulses.  For each random initialization, the control algorithm is used to determine a minimum energy pulse that splits atoms into their first momentum state $|\pm 2\hbar k\rangle$.   Our approach for high order beamsplitters uses the computed solution as the initialization of the control algorithm for beamsplitting to the $|\pm 4\hbar k\rangle$ momentum state.  The process is continued until the desired momentum state $|\pm 2n_0\hbar k\rangle$ is achieved.  In the following, all results are obtained by optimization in dimensionless time $\omega_r t$ over the interval $[0,2\pi]$ using 630 time points. Computations are done in Matlab on a MacBook Pro.

\subsection{Pulse Design Repeatability} \label{sec:convergence}

Convergence of the iterative algorithm is analyzed for the deterministic case in which nominal parameter values $k/k_0=0$ and $\gamma=1$ are fixed.   Figure \ref{fig:error} displays the error at each iteration of the algorithm for target momentum states ranging from $|\pm 2\hbar k\rangle$ to $|\pm 40\hbar k\rangle$.  The realizations in the figure correspond to ten random initializations of the control pulse.  In general, the number of iterations required for the algorithm to converge increases with increasing target momentum spread.  Convergence of the algorithm is consistent across realizations for momentum states up to $|\pm 30 \hbar k\rangle$.  Convergence to higher momentum separations is less consistent primarily because of lower and upper bounds on pulse magnitude that force the control to take bang-bang steps between these bounds.  Pulse magnitude and area are analyzed in the following section.  Figure \ref{fig:error} illustrates that the approach is reliable for the design of high momentum beamsplitters that can be realized with high fidelity.

\begin{figure*}
    \centering
        \includegraphics[width=\linewidth]{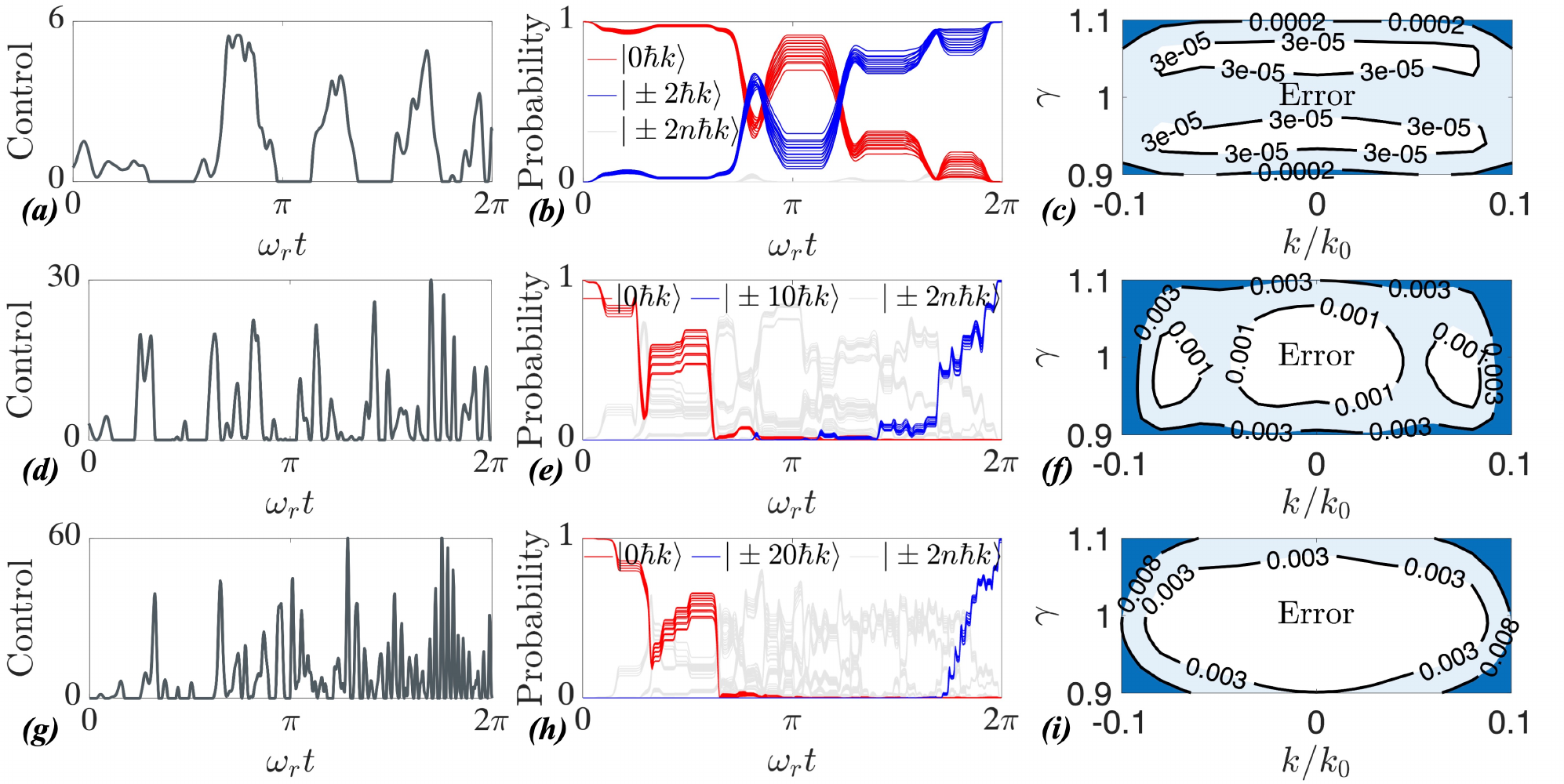}
    \caption{Beamsplitting from $|0\hbar k\rangle$ to $|\pm 2n_0 \hbar k\rangle$, for $n_0=1,\;5,\;10$. Control and probability  refer to the quantities $u/\omega_r$ and $||+2n \hbar k\rangle|^2+||-2n \hbar k\rangle|^2$.  Error refers to the evaluation of $1-||+2n_0 \hbar k\rangle|^2-||-2n_0 \hbar k\rangle|^2$ at $t= T$, i.e., the probability that the momentum state is in energy levels other than the target state $2n_0$.}
    \label{fig:examples}
    \vspace{-2ex}
\end{figure*}

\vspace{-2ex}
\subsection{Pulse Intensity} \label{sec:intensity}

Several metrics are used to evaluate the intensity of a control pulse $u$.  We define the maximum magnitude $MM[u]=\max_t u(t)/\omega_r$, maximum slew rate $MS[u]=\max_t \dot{u}/\omega_r$, area $A[u]=\int_0^Tu(t)dt/\omega_r$, and total variation $TV[u]=\int_0^T|\dot{u}(t)dt/\omega_r$.
These metrics are shown in Figure \ref{fig:energy} for each target momentum state ranging from $|\pm 2\hbar k \rangle$ to $|\pm 40\hbar k \rangle$.  The vertical bars indicate standard deviation across ten random initializations of the iterative algorithm.  Pulse magnitude and area increase almost linearly with increasing target momentum states up to $|\pm 20 \hbar k \rangle$.  In  this range, the standard deviation in pulse magnitude across realizations is negligible, which demonstrates the repeatability of the control algorithm.  For larger momentum states, standard deviation in pulse magnitude and area are more significant because of the amplitude restrictions and resulting bang-bang outcomes, i.e., switching between values that define upper and lower amplitude bounds.  The designed pulses have lower maximum magnitude and area than corresponding pulses in earlier studies \cite{wu2005splitting}.  Table 1 in \cite{wu2005splitting} reports maximum pulse magnitudes ranging from $\max u/\omega_r=2.83$ to $\max u/\omega_r=124$ for target momentum states ranging from $|\pm 2\hbar k\rangle$ to $|\pm 10\hbar k\rangle$.  Figure \ref{fig:energy} shows that the maximum pulse magnitudes computed by the proposed algorithm may be decreased by up to an order of magnitude for a given momentum state.

\begin{table*}[!t]
\caption{\label{tab:poly_data} Compensation using Legendre expansion for design intervals $k/k_0\in [-0.4,0.4]$ and $\gamma \in [0.6,1.4]$.}\vspace{-1ex}
\begin{ruledtabular}
\begin{tabular}{ c|cccc|cccc|cccc  } 
\multicolumn{1}{c}{}& \multicolumn{4}{c}{$|\pm 2\hbar k\rangle$}& \multicolumn{4}{c}{$|\pm 4\hbar k\rangle$}& \multicolumn{4}{c}{$|\pm 6\hbar k\rangle$}  \\
\hline
Poly. Degree/Interval & 1 &  2  &  3  &  4 &   1 &  2  &  3  &  4 &   1 &  2  &  3  &  4 \\
\hline 
$\max \text{ Error}$ & 0.69 & 0.45 & 0.17 & 0.22  &   0.92 & 0.69 & 0.33 & 0.31 &   0.96, & 0.77 & 0.83 & 0.58 \\
$\text{mean Error}$ & 0.24& 0.05& 0.03& 0.02 &   0.42 & 0.17 & 0.09 & 0.08 &   0.47 & 0.31 & 0.32 & 0.25 \\
$\min \text{ Error}$ & 5e-4  &  7e-4  &  1e-3  &  4e-4 &   2e-3 & 1.2e-2 & 1.0e-2 & 1.2e-2 &   3e-3 & 2.6e-2 & 2e-2 & 3.3e-2 \\
$\max u/\omega_r$ & 8.7 & 15.2 & 18.6 & 23.3 &   35.0 & 34.2 & 27.3 & 33.0 &   41.2 & 35.6 & 39.2 & 33.1 \\
$\text{mean}\, u/\omega_r$  & 2.2 & 2.5 & 2.5 & 2.6 &   3.2 & 3.7 & 3.7 & 3.6 &   4.1 & 4.4 & 4.6 & 4.4 \\
$\max |\Delta u|/\omega_r$ & 1.1 & 7.5 & 10.3 & 15.6 &   13.5 & 14.3 & 13.1 & 18.8 &   17.6 & 17.4 & 16.7 & 15.5 \\
$\text{mean}\, |\Delta u|/\omega_r$  & 0.20 & 0.39 & 0.48 & 0.58 &   1.35 & 1.21 & 1.08 & 1.06 &   1.90 & 1.67 & 1.62 & 1.57  \\
Clock (min, sec)  & 0.7 & 4.2 & 11.5 & 23.6 &   4.9 & 7.4 & 19.9 & 37.4 &  5.6 & 9.3 & 22.9 & 55.7
\end{tabular}
\end{ruledtabular}
\end{table*}

\begin{table*}[!t]
\caption{\label{tab:sample_data} Compensation using equidistant sampling for design intervals $k/k_0\in [-0.4,0.4]$ and $\gamma \in [0.6,1.4]$.}\vspace{-1ex}
\begin{ruledtabular}
\begin{tabular}{ c|cccc|cccc|cccc  } 
\multicolumn{1}{c}{}& \multicolumn{4}{c}{$|\pm 2\hbar k\rangle$}& \multicolumn{4}{c}{$|\pm 4\hbar k\rangle$}& \multicolumn{4}{c}{$|\pm 6\hbar k\rangle$}  \\
\hline
No. Samples/Interval & 2 &  3  &  4  &  5 &   2 &  3  &  4  &  5 &   2 &  3  &  4  &  5 \\
\hline 
$\max \text{ Error}$ & 0.94 & 0.75 & 0.61 & 0.34  &  0.99 & 0.90 & 0.73 & 0.87 & 0.97 & 0.96 & 0.96 & 0.94 \\
$\text{mean Error}$ & 0.32 & 0.28 & 0.14 & 0.06  &  0.50 & 0.36 & 0.29 & 0.20 & 0.64 & 0.53 & 0.39 & 0.28 \\
$\min \text{ Error}$ & 1e-9 & 3.6e-3 & 6.7e-3 & 5.8e-3  &  2e-10 & 1.5e-2 & 5.6e-2 & 2.3e-2 & 2e-9 & 1.3e-2 & 2.5e-2 & 2.8e-2 \\
$\max u/\omega_r$ & 10.9 & 15.5 & 18.0 & 18.8  &  27.1 & 22.6, & 21.9 & 24.5 & 37.1 & 38.1 & 36.4 & 35.6 \\
$\text{mean}\, u/\omega_r$  & 1.6 & 2.2 & 2.8 & 2.6  & 3.3 & 4.1 & 4.4 & 5.0 & 4.7 & 4.9 & 4.8 & 4.9 \\
$\max |\Delta u|/\omega_r$ & 1.4 & 4.0 & 4.0 & 4.9  &  9.9 & 8.7 & 8.9 & 9.6 & 11.5 & 14.1 & 15.4 & 17.1 \\
$\text{mean}\, |\Delta u|/\omega_r$ & 0.19 & 0.29 & 0.37 & 0.43  &  0.94 & 0.93 & 0.91 & 1.00 & 1.09 & 1.43 & 1.33 & 1.40\\
Clock (min, sec)  & 0.6 & 3.0 & 6.5 & 14.5  &  3.4 & 5.9 & 9.3 & 23.7 & 1.1 & 5.1 & 13.1 & 35.9
\end{tabular}
\end{ruledtabular}
\end{table*}

\vspace{-2ex}
\subsection{Pulse Bandwidth} \label{sec:filter}

Here we examine the performance of filtered pulses applied to the simulated system evolution.  Implementation of beamsplitting in practice may be subject to low-pass filtering effects inherent in digital control systems, actuator electronics, or optical equipment.  Because the pulses in this study are designed on 629 time intervals over dimensionless time $\omega_r t \in [0,2\pi]$, the step time is $\Delta t = 2\pi/629\omega_r$, where we use $\omega_r = 7.66$ kHz for $^{87}$Rb.  Therefore, the sampling frequency is given by $f_s=629\omega_r/2\pi=3.24$ MHz.  We apply two hundred Butterworth infinite impulse response (IIR) filtering operations to the computed pulses for the targeted momentum states, where each filter indexed by $k$ uses the cutoff frequency $f_c = 0.5f_sk/201$.  The filtered pulses are applied in simulations of the beamsplitting process and the computed probability to populate the target state is shown in Figure \ref{fig:frequency}.  Several conclusions can be drawn from this figure.  First, all target states can be populated with high probability using filtered pulses of sufficiently high bandwidth.   Incidentally, pulse modulation is relatively insensitive to specific times of impulses, lags between impulses, and intensities of impulses comprising the pulse sequence.  Second, the probability of populating the target momentum state increases from zero to unity as the cutoff frequency of the pulse sequence increases from zero to the Nyquist frequency $0.5f_s$.  Third, higher momentum states require higher frequency components, as expected by the bang-bang control limit discussed above.  As verified below, the results of this section suggest that the computed pulses have appropriate properties for realization of atom interferometry experiments.

\section{Robust Beamsplitting Results} \label{sec:results}

In this section, beamsplitting examples that compensate for variability on intervals $k/k_0\in [-0.1,0.1]$ and $\gamma \in [0.9,1.1]$ are first presented to visualize properties of the designed control pulses.  Second, comparison of our robust method to the direct sampling approach is documented for three additional beamsplitting examples that compensate for variability on intervals $k/k_0\in [-0.4,0.4]$ and $\gamma \in [0.6,1.4]$.  Third, the performance of our pulse designs for high momentum state generation is verified in the laboratory.

\begin{figure*}[!t]
    \centering
        \includegraphics[width=.85\linewidth]{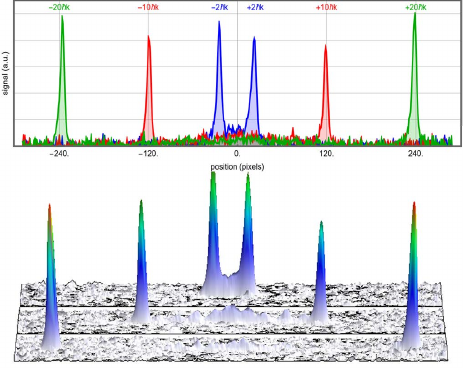}
    \caption{Diffraction from $|0\hbar k\rangle$ into momentum states $|\pm 2 \hbar k\rangle$, $|\pm 10 \hbar k\rangle$, and $|\pm 20 \hbar k\rangle$. }
    \label{fig:experiments}
    \vspace{-2ex}
\end{figure*}

\vspace{-2ex}
\subsection{Simulation} \label{subsec:simulation}

Figure \ref{fig:examples} shows three examples along the top, middle, and bottom rows, respectively, for beamsplitting into target states $|\pm 2 \hbar k\rangle$, $|\pm 10 \hbar k\rangle$, and $|\pm 20 \hbar k\rangle$. Each row displays the control pulse, the evolution of momentum state probabilities, and the terminal error with respect to the desired state. The evolution of probabilities and terminal error realizations are obtained by simulating Eq. \eqref{eq:Raman_Nath} when applying the computed control repeatedly for samples of $k/k_0$ and $\gamma$ within the design intervals. The control pulse $u(\omega_r t)/\omega_r$ is restricted to positive values less than 30 for $|\pm 10 \hbar k\rangle$ and less than 60 for $|\pm 20 \hbar k\rangle$. Observe that the maximum amplitudes of robust controls computed to compensate for parameter variation that are shown in Figure \ref{fig:examples} are greater than the amplitudes of the deterministic counterparts in Figure \ref{fig:error}. Although slightly higher fidelity may be achieved by using higher order polynomial approximation, these three examples use a Legendre expansion truncated up to only first degree polynomials to emphasize the significance of this approach.  Uniform fidelity on the parameter space is generally not achievable with robust optimization using sampled systems of equivalent sizes, which in this case correlates to two samples over both parameter intervals.

Three examples are used to compare robust control in which both parameters may vary in design intervals $k/k_0\in [-0.4,0.4]$ and $\gamma \in [0.6,1.4]$.  Such variations are particularly relevant for atom interferometry experiments in which the cloud of ultra-cold atoms does not necessarily realize Bose-Einstein condensation and that experience significant platform disturbances.  Table \ref{tab:poly_data} shows an analysis for pulse designs that split an atomic sample into momentum states $|\pm 2\hbar k\rangle$, $|\pm 4\hbar k\rangle$, and $|\pm 6\hbar k\rangle$ with variability $k/k_0\in [-0.4,0.4]$ and $\gamma \in [0.6,1.4]$. For each instance, the control algorithm described in Section \ref{sec:control} is applied using Legendre polynomial expansions truncated respectively to first, second, third, and fourth degree polynomials.   Robust pulse designs for these same examples are also synthesized using the alternative approach of uniformly sampling parameter values in the  dynamic constraints, with results in Table \ref{tab:sample_data}. The shorthand $|\Delta u|$ in Tables \ref{tab:poly_data} and \ref{tab:sample_data} is used to refer to the finite-difference derivative approximation of the control signal $u$.  By increasing the degree of Legendre polynomials and the number of samples, the error metrics generally decrease for both methods, but the mean and maximum errors achieved with the proposed Legendre approximation approach decrease at a significantly higher rate than those afforded by direct sampling.  The Legendre expansion provides improved robustness with respect to direct sampling in terms of error metrics and total computational time.  Based on the comprehensive set of computational verification studies described above, we were able to synthesize pulses that are successfully validated in laboratory experiments, described below.

\vspace{-2ex}
\subsection{Experimental Validation} \label{subsec:experiment}

Although the optimized Bragg beamsplitter control pulses described above demonstrate substantial conceptual and computational advances, their importance for atom interferometry lies in their practical realization in experimental hardware. We have carried out preliminary experiments in a guided $^{87}$Rb interferometer using a 780 nm standing wave to provide this validation of our computational method.  The apparatus and methods are described in detail in a recent publication \cite{krzyzanowska2023matter}. After applying the Bragg pulse sequence, we allow the atoms to propagate along the guide for the same elapsed time. In-guide absorption images taken at that time are shown in Fig. \ref{fig:experiments} for synthesis of momentum states $|\pm 2 \hbar k\rangle$, $|\pm 10 \hbar k\rangle$, and $|\pm 20 \hbar k\rangle$, corresponding to the experimental design goals.  The images distinctly show diffraction into the target momentum states, and these results verify that the optimized control pulses can be realized with contemporary laboratory hardware. 

\vspace{2ex}
\section{Conclusions} \label{sec:conc}
\vspace{-1ex}

We have presented an optimal control synthesis technique that enables a new level of control over matter-wave dynamics in atom interferometry. Principles from optimal quantum control and polynomial approximation are incorporated to yield Bragg beamsplitter control designs that surpass previously demonstrated fidelity and momentum level performance, and that also provide practical guarantees of insensitivity to variation in system parameters and environmental noise. The approach achieves both unprecedented momentum transfer and provable robustness to amplitude, frequency, and detuning fluctuations, which are key advances for the transition of atom interferometers from laboratory demonstrations to field-deployable quantum sensors. Critically, our experimental verification of the designed control pulse performance confirms that advanced optimal control methods are practical for manipulating the nonlinear dynamics of Bose–Einstein condensates.

Our rapid numerical optimization method enables the synthesis of robust control pulses that are compatible with laboratory pulse modulation systems. Moreover, the use of polynomial expansion is shown to have clear advantages in compensating for parameter variability over sampling-based approaches in terms of specific performance metrics.  In the exposition of our control synthesis approach, we distinguish between two separable Hilbert spaces $\mathcal H$ and $\mathcal L$, which are defined for the momentum of the quantum system and the domain of parameter variability in the Hamiltonian, respectively. The numerical representation requires careful disambiguation of the bases and expansions used in the joint space. We emphasize that our control synthesis approach is a constructive and is implemented in a public source code repository \cite{baker2025beast}, and is now validated in the laboratory.

While we examine multi-photon Bragg scattering in atom interferometry in the specific case of symmetric Bragg diffraction with a standing wave, there is no impediment to extension of the symmetric formulation here to the non-symmetric setting.  It remains an open question what non-symmetric states are reachable starting from a perfectly symmetric initial state, because the symmetry will remain preserved under all evolution with the control parameters available in the model used here.  The algorithm could be modified to start with a prepared state that has some initial momentum/phase asymmetry.  This motivates a study of the controllability of Bragg beamsplitter systems in general.  As demonstrated in our experiment, because we are defining the terminal error to be minimized as $1-||+2n_0 \hbar k\rangle|^2-||-2n_0 \hbar k\rangle|^2$ at $t= T$, the absolute phase of the two components of the target state ($|-2nhk\rangle$ and $|+2nhk\rangle$) is not enforced.  However, because of the symmetry in the ensemble dynamics, the relative phase is always antisymmetric. Enforcing terminal state phase directly as an optimization constraint would make the pulse synthesis algorithm far more restrictive and result in slower numerical convergence, so that leaving absolute phase as a degree of freedom is a key factor that enables the algorithm to converge rapidly.  The experimental results verify that our pulses are indeed successful in atom interferometry at high momentum levels , i.e., $|\pm 20\hbar k\rangle$.

Our synthesis approach could in principle be extended to produce non-symmetric states or more complex entangled states in networks of interacting spin systems.  Promising directions for follow-on studies also include minimum-time controllability and reducing dead-times in the synthesized controls.  Extending our technique to multi-particle entangled states and novel interferometric geometries may allow quantum sensors to surpass the standard quantum limit and offer enhanced sensitivity for tests of fundamental physics, gravitational measurements, and inertial navigation. More broadly, the principles demonstrated here are applicable across diverse platforms in atomic, molecular, and solid-state physics, in which high-fidelity control of quantum states is challenged by parameter variation.   Our concurrent advancement of the computational foundations and experimental implementation of Bragg beamsplitting brings the use of noise-resilient quantum control closer to use in deployable quantum technologies.

\vspace{1ex}
\section*{acknowledgments}
\vspace{-2ex}

This project was supported by the LDRD program and the Center for Nonlinear Studies at Los Alamos National Laboratory.  Research conducted at Los Alamos National Laboratory is done under the auspices of the National Nuclear Security Administration of the U.S. Department of Energy under Contract No. 89233218CNA000001. Report No.: LA-UR-25-32246.

\bibliography{references}

\end{document}